\documentclass[%
 reprint,
superscriptaddress,
 amsmath,amssymb,
 aps,
]{revtex4-2}

\usepackage{graphicx}
\usepackage{dcolumn}
\usepackage{bm}
\usepackage[dvipsnames]{xcolor}
\usepackage{soul}
\usepackage{afterpage}
\usepackage{hyperref}
\usepackage{xr}
\usepackage{amsmath}
\makeatletter

\usepackage{gensymb}

\def\maketitle{
\@author@finish
\title@column\titleblock@produce
\suppressfloats[t]}

\makeatother


\begin{document}

\title{Understanding Damping Mechanisms via Spin Diffusion Length in Low-damping Li$_{0.5}$Al$_{1.0}$Fe$_{1.5}$O$_4$ Spinel Ferrite Thin Films}

\author{Katya Mikhailova}
\affiliation{Department of Applied Physics, Stanford University, Stanford, CA, 94305, USA}
\affiliation{Geballe Laboratory for Advanced Materials, Stanford University, Stanford, California 94305, USA}
\affiliation{Stanford Institute for Materials and Energy Sciences, SLAC National Accelerator Laboratory, Menlo Park, California 94025, USA}

\author{Lerato Takana}
\affiliation{Department of Applied Physics, Stanford University, Stanford, CA, 94305, USA}
\affiliation{Geballe Laboratory for Advanced Materials, Stanford University, Stanford, California 94305, USA}

\author{Guanxiong Qu}
\affiliation{Department of Physics and Astronomy, University of California, Irvine, Irvine, CA 92697, USA}

\author{Juan A. Hofer}
\affiliation{Department of Physics and Center for
Advanced Nanoscience, University of California San Diego, La Jolla, CA 92093, USA}

\author{Herv{\'e} M. Carruzzo}
\affiliation{Department of Physics and Astronomy, University of California, Irvine, Irvine, CA 92697, USA}

\author{Ivan K. Schuller}
\affiliation{Department of Physics and Center for
Advanced Nanoscience, University of California San Diego, La Jolla, CA 92093, USA}

\author{Clare C. Yu}
\affiliation{Department of Physics and Astronomy, University of California, Irvine, Irvine, CA 92697, USA}

\author{Yuri Suzuki}
\affiliation{Department of Applied Physics, Stanford University, Stanford, CA, 94305, USA}
\affiliation{Geballe Laboratory for Advanced Materials, Stanford University, Stanford, California 94305, USA}
\affiliation{Stanford Institute for Materials and Energy Sciences, SLAC National Accelerator Laboratory, Menlo Park, California 94025, USA}

\date{\today}

\begin{abstract}
The mechanisms underlying magnon damping are of fundamental and technological interest in low-damping materials. We find low-damping ferrimagnetic insulator Li$_{0.5}$Al$_{1.0}$Fe$_{1.5}$O$_{4}$ (LAFO) thin films to be a promising model system for probing these mechanisms because of its distinct temperature dependent spin diffusion length (SDL) trends for electrically and thermally generated magnons. With increasing temperature, the electrical SDL shows minimal change, while the thermal SDL decreases. We attribute these trends to distinct magnon populations and scattering mechanisms: thermally generated high $k$ magnons are limited by magnon-phonon scattering, whereas electrically generated low $k$ magnons are limited by relaxational scattering from magnetic impurities. 
\end{abstract}

\maketitle

\textit{Introduction.---} 
Magnons, the quanta of collective spin excitations, and their transport have been an increased focus of recent research, as they may act as high-speed, energy-efficient information carriers for next-generation computing and quantum technologies \cite{chumak_2015}. A central challenge is identifying and controlling the mechanisms governing magnon propagation length and lifetime, and understanding their dependence on the excitation mechanism. Magnons can be generated by injecting angular momentum into a system through thermal gradients, microwave excitations, spin-transfer and spin-orbit torques, or optical excitations \cite{demokritov_2012}. Understanding their associated damping mechanisms is an important step towards spin-wave-based spintronics in ferrimagnetic insulators (FMIs), where the absence of charge currents reduces power dissipation and improves energy efficiency \cite{chumak_2015}.

Magnon propagation has been characterized and quantified via the spin diffusion length (SDL), the characteristic magnon propagation length. Nonlocal transport measurements in magnetic insulator films, using magnon injector and detector electrodes, have been employed to quantify magnon transport and extract the SDL \cite{cornelissen_2015,Goennen_2015,Shan_2016,shan_2017,Giles_2017,gray_2018,riddiford_2019,Oyanagi_2020,li_2022,Wei_2022,gao_2022, Li_2023}. In these experiments, magnons are generated electrically through spin transfer torque and thermally via the spin Seebeck effect (SSE), and detected via the inverse spin Hall effect (ISHE) \cite{cornelissen_2015}. The SDL and its temperature dependence provide key insight into the underlying damping mechanisms. Previous studies in the gold-standard FMI Y$_3$Fe$_5$O$_{12}$ (YIG) show that thermally and electrically generated magnons exhibit a similar temperature dependence, suggesting common damping mechanisms \cite{cornelissen_2016,gomez-perez_2020}. However, the origin of the non-monotonic temperature dependence of the SDL remains unresolved, and it is unclear whether different generation mechanisms probe the same damping channels or whether nonlocal transport can distinguish them.

We investigate magnon transport in Li$_{0.5}$Al$_{1.0}$Fe$_{1.5}$O$_4$ (LAFO) thin films, a promising FMI material for spintronic applications \cite{zheng_2020, zheng2023, Omahoney2023, ren_2023, alaei_2025,takana_2025,tong_2026}. 
Compared to YIG, LAFO films have comparable ultra-low magnetic damping ($\alpha\sim10^{-4}$), a simpler crystal structure, epitaxial growth without interfacial layers, and strain-tuned magnetic anisotropy \cite{sun_2012,onbasli_2014,soumah_2018,cooper_2017,suturin_2018,gaur_2023}. Moreover, oxide spinels provide a pathway toward integration with conventional silicon-based platforms: unlike YIG, LAFO films can be grown on oxide perovskites that can be integrated with silicon and benefit from a lower growth temperature \cite{arsad_2023,mori_1991,kang_2006,baek_2013}. These properties make LAFO an attractive system for probing magnon transport mechanisms and the associated SDLs. 

In this paper, we demonstrate experimentally and explain theoretically that magnon propagation in LAFO is governed by distinct scattering processes, with thermally and electrically generated magnons dominated by different mechanisms. At room temperature, the SDLs for thermally and electrically generated magnons are similar and consistent with values reported for other sub-100 nm thick spinel and garnet FMI thin films \cite{shan_2017,li_2022,Wei_2022,gao_2022}. However, they diverge at lower temperatures, a trend not observed in YIG. We attribute this divergence to distinct dominant scattering mechanisms. Thermally generated magnons are characterized by large $k$ magnons, are primarily limited by magnon-phonon and Rayleigh scattering, whereas electrically generated magnons,  characterized by low $k$ magnons, are limited by magnetic impurity scattering, 
modeled as two-level systems (TLS)~\cite{VVleck1963,Seiden1964}. This direct experimental separation of damping channels within a single material establishes a quantitative link between the magnon generation mechanism, momentum distribution, and dominant scattering process, and provides a general framework for probing and engineering magnon transport in FMIs.


\afterpage{
\begin{figure}[!htbp]
    {\includegraphics[width=0.93\columnwidth]{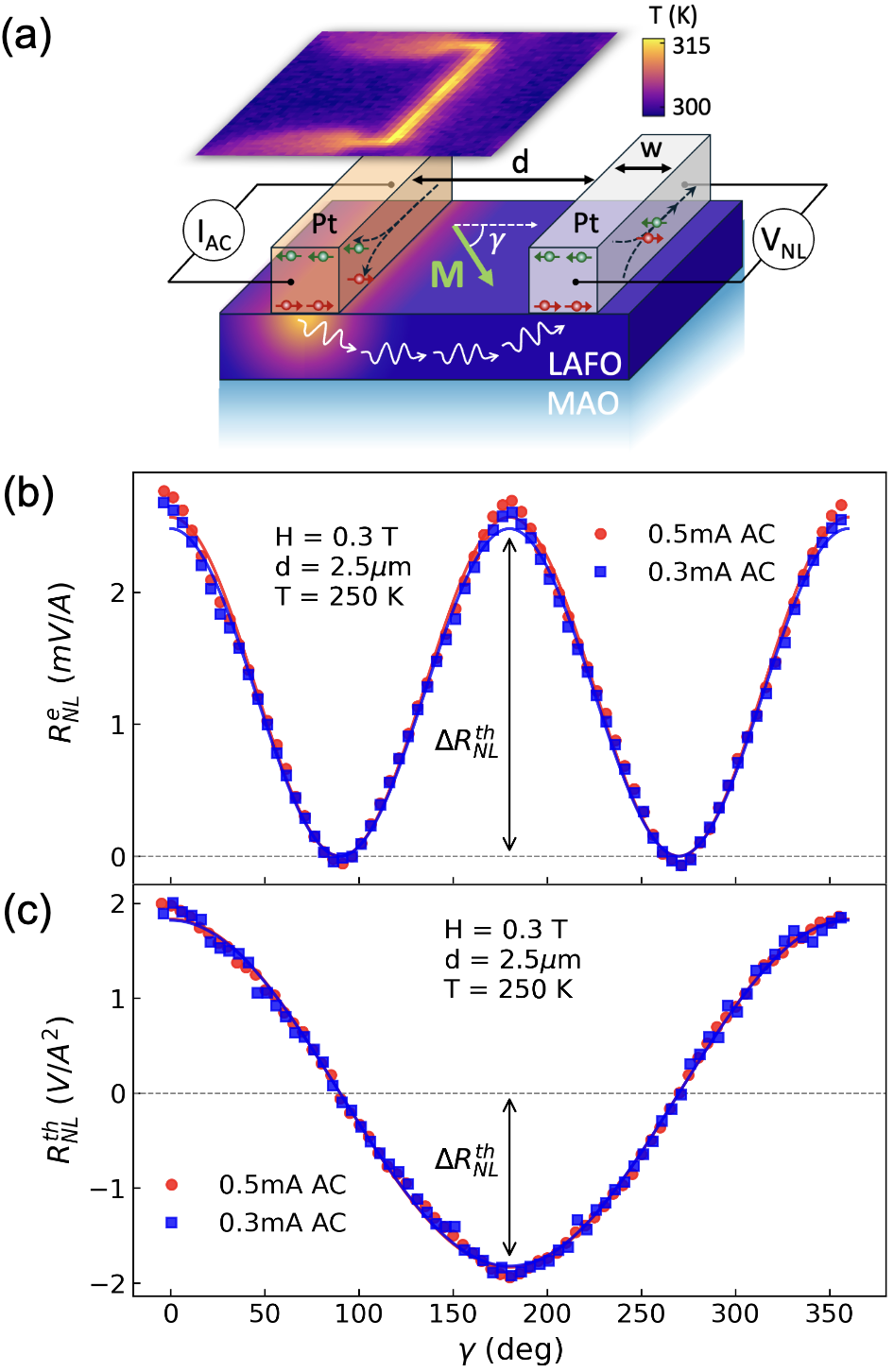}}
    \caption{Nonlocal transport experimental setup. (a) Schematic of a device consisting of Pt injector and detector strips (width $w$) on a LAFO thin film grown on MgAl$_2$O$_4$ (MAO). Magnons are injected into LAFO electrically via the spin Hall effect (SHE) and thermally via the spin Seebeck effect, and detected a distance $d$ away via the inverse SHE. The LAFO magnetization $M$ is rotated in-plane by an angle $\gamma$ using an in-plane external magnetic field $H$. The image above the schematic is a mid-wave infrared (MWIR) microscopy image taken at $T = $ 300 K for a device with $d=5.0\,\mu$m when a 0.5 mA ac current is applied. (b, c) Nonlocal resistance (defined from harmonic voltages) as a function of in-plane angle $\gamma$ in a device with $d=2.5\,\,\mu$m with $H=0.3$ T for two ac currents: 0.3 mA (blue) and 0.5 mA (red). (b) Electrical nonlocal resistance with fits to Eq.~\eqref{nonloceq1} where $\Delta R_{NL}^{e}$ is the electrical nonlocal amplitude. (c) Thermal nonlocal resistance with fits to Eq.~\eqref{nonloceq2}, where $\Delta R_{NL}^{th}$ defines the thermal nonlocal amplitude. }
    \label{fig:setup}
\end{figure}
}

\textit{Results.---}
We extract the SDL in LAFO thin films using nonlocal transport measurements (Fig.~\ref{fig:setup}(a)) on 16 nm and 86 nm thick films grown on single-crystalline MgAl$_2$O$_4$ (MAO) substrates by pulsed laser deposition (See Supplemental Section \ref{sup:sample prep} \cite{supp}). The device geometry consists of two Pt electrodes (width w = 0.2 $\mu$m, thickness 4 nm) separated by a distance $d$, serving as injector and detector. Magnons are generated at the injector via spin accumulation from the spin Hall effect (SHE) and via Joule heating-induced thermal gradients. Magnons are detected by the detector through the inverse spin Hall effect (ISHE). The electrically and thermally generated magnon contributions are distinguished using harmonic detection. An ac current $I_{AC} = I_0 \sin(\omega t)$ ($\omega/(2\pi)\sim9.8$ Hz) was applied to the injector. The resulting spin accumulation at the Pt/LAFO interface generates magnons in LAFO via exchange coupling, producing a first harmonic nonlocal voltage $V_{NL}^{1\omega}\propto I_{AC}$ \cite{chumakDirectDetectionMagnon2012a}. Simultaneously, Joule heating generates a temperature gradient that drives magnons via the spin Seebeck effect (SSE), producing a second harmonic nonlocal voltage $V_{NL}^{2\omega} \propto I_{AC}^2$ \cite{uchida_2010}. 

To separate the electrical and thermal contributions, we perform in-plane magnetic field rotation by an angle $\gamma$ (Fig.~\ref{fig:setup}(a)). The spin pumping efficiency depends on the relative orientation between the LAFO magnetization $M$ and the Pt spin accumulation.
The first harmonic signal has a $\cos^2(\gamma)$ dependence as both the injection and detection processes depend on the angle. The second harmonic has a $\cos(\gamma)$ dependence, as only the detection process depends on the angle. The electrical and thermal nonlocal resistances are defined as 
\begin{align}
        R^{e}_{NL} &= V^{1\omega}_{NL}/I_0 = \Delta R^{e}_{NL} \cos^2(\gamma), \label{nonloceq1} \\
        R^{th}_{NL} &= V^{2\omega}_{NL}/I_0^2 = \Delta R^{th}_{NL} \cos(\gamma). \label{nonloceq2}
\end{align}
where $V^{1\omega}_{NL}$ and $V^{2\omega}_{NL}$ are the lock-in first and second harmonic voltages, $I_0$ is the injector ac amplitude, and $\Delta R^e_{NL}$ and $\Delta R^{th}_{NL}$ are the electrical and thermal nonlocal resistance amplitudes. Figure \ref{fig:setup}(b, c) shows the angular scans at 250 K for two currents ($I_0=0.3$ mA and 0.5 mA) with an applied field $H= 0.3$ T, sufficient to fully saturate the LAFO magnetization (see Supplemental Section \ref{sup:characterization} \cite{supp}). The resistance data follow the expected angular dependence as given by Eq.~\eqref{nonloceq1} and Eq.~\eqref{nonloceq2} and are independent of current, because the voltage signals are normalized by their respective currents.

\afterpage{
\begin{figure}[!htbp]
   \centering
   {\includegraphics[width=0.9\columnwidth]{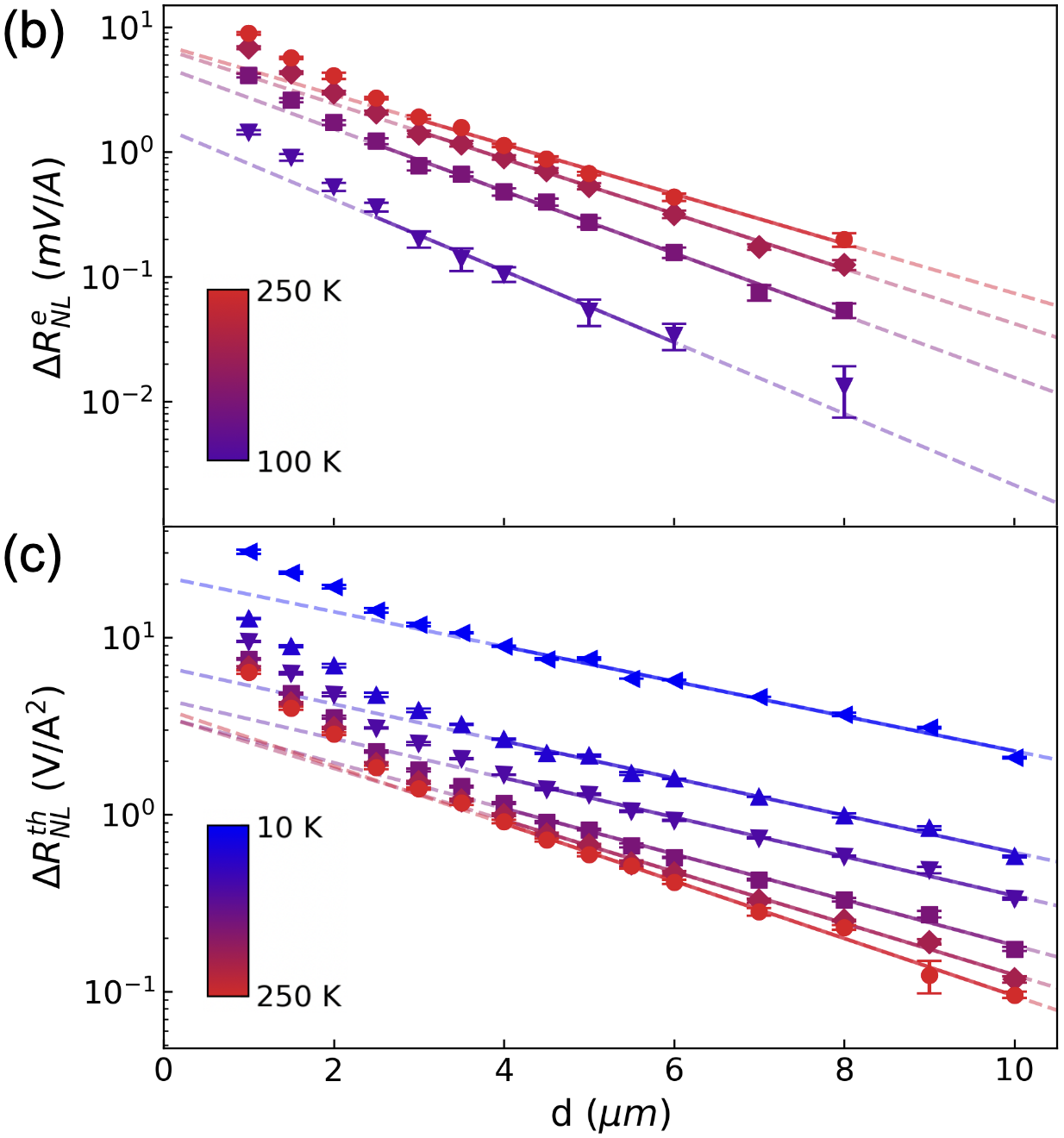}}
   \caption{Nonlocal resistance amplitude as a function of injector-detector distance $d$ for (a) electrically and (b) thermally generated magnons for select temperatures. Solid lines are fits to Eq.~\eqref{eq:NL exp sol} and are used to extract $\lambda_m^{e,th}$, and dashed lines are extrapolations beyond the fitted range. The data are well described by a single exponential decay over the selected distance range. 
   }
   \label{fig:data}
\end{figure}
}


We measured the nonlocal resistance amplitude as a function of injector-detector separation $d$ at different temperatures and extracted $\lambda_m^{e,th}$ from exponential fits. The distance dependence of the nonlocal resistance, along with fits, for electrically and thermally generated magnons is shown in Fig. \ref{fig:data}. Magnon transport in thin films is described by one-dimensional magnon diffusion \cite{cornelissen_2015}, yielding an exponential decay of the magnon density with characteristic length $\lambda_m^{e,th}$ (see Supplemental Section \ref{sup:SDL equation derivation} \cite{supp}). In the regime $ d > \lambda_m$, the nonlocal resistance is well described by
\begin{equation}
   \Delta R_{NL}^{e,th}(d) = \frac{C}{\lambda_m^{e,th}}e^{-d/\lambda_m^{e,th}},
   \label{eq:NL exp sol}
\end{equation}
and $\lambda_m^{e,th}$ is extracted by fitting to $\ln(\Delta R_{NL}^{e,th})$, which weights data at all distances more equally. As seen in Fig. \ref{fig:data}, the data are well described by the exponential function over the selected fitting range, with deviations from the exponential fit occurring at small $d$, where magnon transport is in the diffusive regime \cite{cornelissen_2017} and thermal gradients in the LAFO film remain significant.

To ensure reliable extraction of $\lambda_m^{e,th}$, we therefore restrict the fitting range to the exponential regime. The thermal decay length of the injector-induced temperature profile is determined using mid-wave infrared (MWIR) imaging \cite{hofer_2025} (shown above the schematic in Fig.~\ref{fig:setup}(a), yielding a decay length $\ell_{th} \sim 2.4 \,\mu$m (See Supplemental Section \ref{sup:thermal} \cite{supp}). This sets the lower bounds of the fitting ranges: $d \ge 2.5\,\mu$m for electrical magnons and $d \ge 4 \,\mu$m for thermal magnons. We use a larger lower bound for the thermal SDL fit to avoid as much of the region with a temperature increase as possible. For electrical magnons, we additionally exclude long-distance data ($d > 8\,\mu$m) and $T<90$K data, where the signal approaches the noise floor (See Supplemental Section \ref{sup:RvsT diff devs} \cite{supp}). The solid lines in Fig. \ref{fig:data} show the exponential fits, and the dashed lines indicate extrapolations beyond the fitting range.

From these fits, we extract the SDL and find starkly different temperature dependences for electrically and thermally generated magnons (Fig. \ref{fig:SDL vs T w/ theory}). We find that thermally generated magnons exhibit a decreasing SDL from $\sim 4.4 \,\mu$m at 10 K to $\sim 2.6 \,\mu$m at 280 K. Electrically generated magnons show the opposite trend, increasing from $\sim 1.3 \,\mu$m at 90 K to $\sim 2.4 \,\mu$m at 280 K. These opposite trends highlight a fundamental difference between thermally and electrically generated magnons in LAFO.


To explain the distinct temperature dependences of the SDL, we model magnon diffusion~\cite{cornelissen_2015} in terms of two nonequilibrium magnon populations with different momentum distributions. Electrically generated magnons are assumed to be dominated by small-$k$ modes, whereas thermally generated magnons are dominated by large-$k$ modes. Accordingly, we introduce a cutoff energy $\hbar\omega_{\mathrm{cut}}=5.3~\mathrm{K}$, which serves as the upper bound of the electrical magnons and the lower bound of the thermal magnons. Electrical magnons are injected locally at the Pt/LAFO interface through interfacial $s$-$d$ coupling~\cite{zhang_2012Spin,Bender_2015}. Because spin transfer is most efficient when the magnon energy matches the Pt spin accumulation, $\mu_s \approx 80~\mu\mathrm{eV}$ \cite{supp}, these magnons predominantly occupy the low-energy sector, with a characteristic energy $\hbar\omega_e \approx 0.93~\mathrm{K}$. Although this initial distribution may broaden through magnon-magnon and magnon-phonon scattering, both processes become less effective in the long-wavelength limit~\cite{Dyson_1956}, allowing the low-energy character of electrically injected magnons to persist. In contrast, thermally generated magnons arise nonlocally from the bulk spin Seebeck effect near the injector~\cite{Shan_2016,cornelissen_2017} and are therefore expected to exhibit a much broader momentum distribution. This distinction in nonequilibrium magnon populations naturally leads to their opposite transport behaviors.

\afterpage{
\begin{figure}[!htbp]
    \centering
    {\includegraphics[width=0.93\columnwidth]{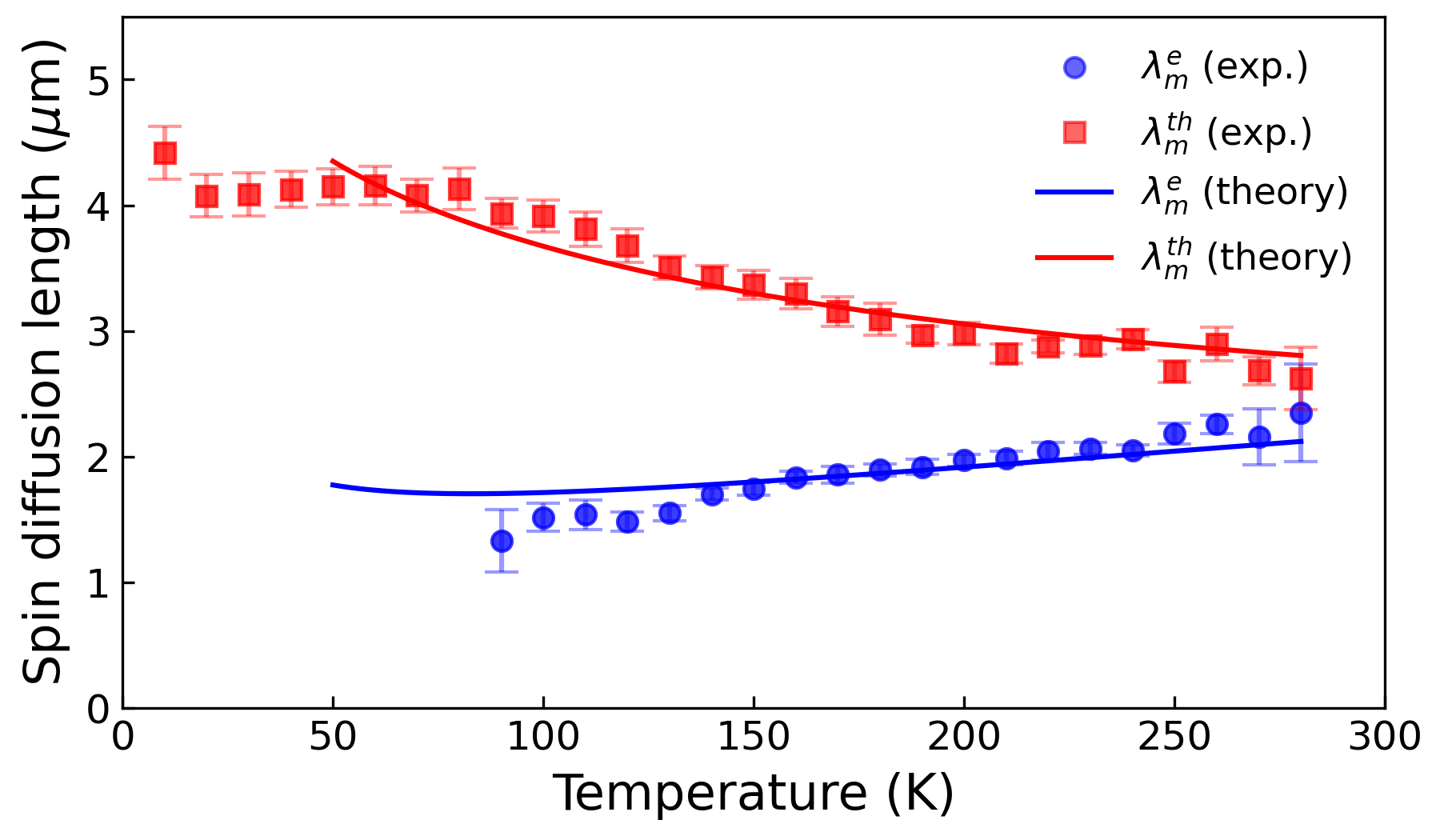}}
    \caption{Spin diffusion length as a function of temperature for thermally ($\lambda_m^{th}$) and electrically generated magnons ($\lambda_m^{e}$). Circles denote the experimental data extracted from fits to the nonlocal resistance using Eq.~\eqref{eq:NL exp sol}, while solid lines denote the theoretical modeling. The parameters for the model are listed in Supplemental Section \ref{sup:theory} \cite{supp}. 
    }
    \label{fig:SDL vs T w/ theory}
\end{figure}
}

Our model quantitatively reproduces the measured temperature dependence of the SDL down to $50~\mathrm{K}$ (see Fig.~\ref{fig:SDL vs T w/ theory} and Supplemental Section \ref{sup:theory} \cite{supp}). The SDL is given by $\sqrt{D\tau_{\mathrm{nc}}}$, where 
both the diffusion constant $D=\langle v^2\tau_{\mathrm{c}}\rangle$ and magnon-number-nonconserving relaxation time $\tau_{\mathrm{nc}}$ are averaged over the corresponding nonequilibrium magnon distributions. 
Magnon relaxation is decomposed into magnon-number-conserving $\tau_{\mathrm{c}}$ and magnon-number-nonconserving channels $\tau_{\mathrm{nc}}$, since at high temperature we have $\tau_{\mathrm{c}} \ll \tau_{\mathrm{nc}}$ which allows the magnon system to establish local equilibrium~\cite{cornelissen_2016_2}. Two competing scattering channels determine the temperature dependence of the SDL. The first is magnon-number-conserving $(\propto \tau_{\mathrm{c}})$ and includes Rayleigh scattering from point defects, which scales as $k^4$~\cite{Pekeris1947}, and magnon-phonon scattering, whose rate scales as $k^2$~\cite{Liu2017} and increases approximately linearly with temperature because of the increasing phonon population~\cite{Streib2019}. Acting alone, these processes would reduce the SDL with increasing temperature. Note that at low temperatures ($T<50~\mathrm{K}$), the magnon-phonon relaxation time diverges and other magnon-number-conserving channels, such as elastic two-magnon scattering, are expected to dominate~\cite{cornelissen_2016_2}. We therefore focus on $T>50~\mathrm{K}$, where magnon-phonon scattering remains the dominant magnon-conserving process.

The second channel is magnon-number-nonconserving scattering $(\propto\tau_{\mathrm{nc}})$ from magnetic impurities, modeled as TLS exchange-coupled to the LAFO host spins~\cite{Pal_2026}. In relaxational TLS scattering~\cite{VVleck1963,Seiden1964,Jackle1972}, magnons modulate the TLS level splitting and dissipate energy through the ensuing population readjustment; this channel weakens with increasing temperature as the TLS approach saturation. The opposite temperature dependences of $\lambda_m^{\mathrm{th}}$ and $\lambda_m^{e}$ then follow naturally from their distinct momentum distributions: thermally generated magnons, which occupy larger $k$, are governed primarily by magnon-conserving $k$-dependent scattering, causing $\lambda_m^{\mathrm{th}}$ to decrease with increasing temperature, whereas electrically generated magnons, which occupy smaller $k$, are less affected by these processes and instead are dominated by TLS relaxational scattering, causing $\lambda_m^{e}$ to increase with increasing temperature.



\textit{Discussion.---}
The extraction of $\lambda_m^{e,th}$ allows us to directly compare intrinsic magnon transport behavior across temperature and film thickness. We note that the temperature dependence of the SDL in LAFO is qualitatively different from that reported for YIG films, where $\lambda_m^{th}$ generally increases with decreasing temperature, whereas $\lambda_m^{e}$ sometimes also increases with decreasing temperature or shows minimal change \cite{cornelissen_2016, gomez-perez_2020}. To ensure that this difference does not arise from fitting procedures, we compared SDLs extracted using both the full diffusion solution and the exponential approximation with restricted fitting ranges (see Supplemental Section \ref{sup:SDL fit comparison} \cite{supp}). The extracted SDL values and temperature trends remain unchanged, indicating that the observed trends are not an artifact of the fitting procedure.

The differences in temperature dependence of the SDLs in LAFO and YIG films cannot be attributed to differences in sample thickness. We repeated our measurements on 86 nm thick LAFO films in addition to the 16 nm thick samples (see Supplemental Section \ref{sup:86 nm LAFO data} \cite{supp}). Qualitatively, the trends are the same. The electrical SDL is independent of thickness, while the thermal SDL increases in magnitude for thicker films but retains the same temperature dependence. This indicates that the diverging temperature trends are intrinsic to LAFO and not sample or thickness dependent. 

The increase in the thermal SDL for thicker films is consistent with reduced surface scattering from an increase in the ratio of the magnons traveling through the bulk versus those scattering from the surface of the film. Defects on the surface produce Rayleigh scattering that scales as $k^4$. Since thermally generated magnons are characterized by large wave vectors, they are particularly sensitive to surface scattering. Thicker films have relatively less surface scattering, and longer thermal SDLs are observed. In contrast, electrically generated magnons, which are dominated by small wave vectors, are less sensitive to surface scattering, resulting in less thickness dependence. These thickness dependence trends further support the model assumption that thermal magnons occupy high $k$ states, while the electrical magnons occupy low $k$, providing experimental support for the distinct momentum distributions.

YIG films show temperature dependent SDL trends that differ from what we observe in LAFO. These differences likely arise from the underlying differences in damping mechanisms attributed to magnons observed in LAFO and YIG films. In LAFO, aluminum doping introduces a high density of TLS defects. In this case, TLS scattering can dominate the magnon-nonconserving relaxation processes, leading to a substantial difference in $\tau_{nc}$ and, consequently, in the SDL. In contrast, the smaller difference between $\lambda_m^{e}$ and $\lambda_m^{th}$ in YIG samples~\cite{cornelissen_2016,gomez-perez_2020} suggests nearly identical nonequilibrium magnon distributions for electrically and thermally generated magnons, or a weaker role of TLS scattering in YIG samples, reducing the sensitivity of magnon transport to their distributional differences. Further exploration of the detailed scattering mechanisms at work in garnet versus spinel ferrites is necessary. Probing the nonequilibrium magnon distributions would require momentum-resolved spectroscopy, such as Brillouin light scattering (BLS) or resonant inelastic x-ray scattering (RIXS), with experimental setups that isolate or independently probe electrically and thermally generated magnons.

\textit{Summary.---}
We demonstrate that thermally and electrically generated magnons in LAFO exhibit distinct temperature dependent SDLs, reflecting different underlying damping mechanisms. Thermally generated magnons are attributed to larger $k$ magnons dominated by Rayleigh and magnon-phonon scattering, while electrically generated magnons are attributed to smaller $k$ magnons dominated by magnon-nonconserving relaxational scattering from TLS, consistent with both their SDL temperature and thickness dependence. Our findings reveal the key damping mechanisms for spin diffusion in thermally versus electrically generated magnons, establishing magnon distribution, boundary scattering, and impurity-driven relaxation as key factors governing magnon transport.

\medskip
\textit{Acknowledgments.---}
This work was supported as part of the Center for Energy Efficient Magnonics, an Energy Frontier Research Center (CEEMag) funded by the U.S. Department of Energy, Office of Science, Basic Energy Sciences at SLAC National Laboratory under contract \# DE-AC02-76SF00515. Part of this work was performed at nano@stanford facilities RRID:SCR\_026695. Research at UCSD was supported as part of the Quantum Materials for Energy Efficient Neuromorphic Computing (Q-MEEN-C), an Energy Frontier Research Center funded by the U.S. Department of Energy, Office of Science, Basic Energy Sciences under Award \# DE-SC0019273.

\medskip
\textit{Data availability.---}
The data that support the findings of this article are openly available in the Stanford Digital Repository at \url{https://purl.stanford.edu/fm621hf7395}.

\bibliography{ref}

\clearpage
\onecolumngrid

\title{Supplementary Information for Understanding Damping Mechanisms via Spin Diffusion Length in Low-damping Li$_{0.5}$Al$_{1.0}$Fe$_{1.5}$O$_4$ Spinel Ferrite Thin Films}

\author{Katya Mikhailova}
\affiliation{Department of Applied Physics, Stanford University, Stanford, CA, 94305, USA}
\affiliation{Geballe Laboratory for Advanced Materials, Stanford University, Stanford, California 94305, USA}
\affiliation{Stanford Institute for Materials and Energy Sciences, SLAC National Accelerator Laboratory, Menlo Park, California 94025, USA}

\author{Lerato Takana}
\affiliation{Department of Applied Physics, Stanford University, Stanford, CA, 94305, USA}
\affiliation{Geballe Laboratory for Advanced Materials, Stanford University, Stanford, California 94305, USA}

\author{Guanxiong Qu}
\affiliation{Department of Physics and Astronomy, University of California, Irvine, Irvine, CA 92697, USA}

\author{Juan A. Hofer}
\affiliation{Department of Physics and Center for
Advanced Nanoscience, University of California San Diego, La Jolla, CA 92093, USA}

\author{Herv{\'e} M. Carruzzo}
\affiliation{Department of Physics and Astronomy, University of California, Irvine, Irvine, CA 92697, USA}

\author{Ivan K. Schuller}
\affiliation{Department of Physics and Center for
Advanced Nanoscience, University of California San Diego, La Jolla, CA 92093, USA}

\author{Clare C. Yu}
\affiliation{Department of Physics and Astronomy, University of California, Irvine, Irvine, CA 92697, USA}

\author{Yuri Suzuki}
\affiliation{Department of Applied Physics, Stanford University, Stanford, CA, 94305, USA}
\affiliation{Geballe Laboratory for Advanced Materials, Stanford University, Stanford, California 94305, USA}
\affiliation{Stanford Institute for Materials and Energy Sciences, SLAC National Accelerator Laboratory, Menlo Park, California 94025, USA}
\
\date{\today}

\maketitle

\onecolumngrid
\setcounter{figure}{0}
\renewcommand{\thefigure}{S\arabic{figure}}

\section{Sample Preparation and Measurement Methods}
\label{sup:sample prep}
\subsection{LAFO thin film growth} 
The Li$_{0.5}$Al$_{1}$Fe$_{1.5}$O$_{4}$ (LAFO) thin films were epitaxially grown on single-crystalline MgAl$_{2}$O$_{4}$ (MAO) substrates using pulsed laser deposition (PLD) with a 248 nm KrF laser with a laser fluence of 2.8 J/cm$^2$. The MAO substrate was chosen to induce easy-plane magnetism in the LAFO thin films from the compressive strain induced by the lattice mismatch between LAFO and MAO \cite{Omahoney2023}. 
The MAO substrate was sonicated in acetone, followed by IPA for 5 minutes each. A polycrystalline pressed target of Li$_{0.6}$Al$_{1}$Fe$_{1.5}$O$_{4}$ (produced by Toshiba Corporation) was used. The target contains additional Li to compensate for the Li loss during deposition. Before deposition, the target was pre-ablated for 1 minute at 1 Hz followed by 1.5 minutes at 5 Hz. For deposition, the substrate was heated to a temperature of 425$\degree$C in 15 mTorr of O$_{2}$ and the target was ablated with a repetition rate of 2 Hz. Following the deposition time, the sample was cooled to room temperature in 100 Torr of O$_{2}$. Each LAFO sample growth included two samples: one for the experiment and one for sample characterization.

\subsection{Device fabrication} 
A series of 4 nm platinum two-electrode devices (width 0.2 $\mu$m, length 40 $\mu$m) with a range of separation distances, $d$, from 1 $\mu$m to 12 $\mu$m were patterned on top of the LAFO thin films using electron-beam lithography, Pt sputtering, and lift-off. The electron-beam lithography was performed using the Raith EBPG 5200+ electron beam lithography system on the LAFO film spin coated with a 950 PMMA A2 e-beam resist layer and an Electra 92 conductive layer. After exposure, the pattern was developed using a 1:3 ratio of methyl isobutyl ketone (MIBK) and isopropyl alcohol (IPA). The Pt was dc-sputtered in 3 mTorr of Ar at 200 W, followed by a lift-off procedure using sonication in acetone and IPA. The sample was soaked in acetone for 8 minutes, sonicated for 2 minutes, sonicated in fresh acetone for 5 minutes, and finally, sonicated in IPA for 5 minutes. The 40 nm thick AuPd bonding pads were patterned using optical lithography, AuPd sputtering in 3 mTorr of Ar, and the same lift-off procedure as Pt.

\subsection{Nonlocal transport measurements} 
The nonlocal transport experiments were conducted in a Quantum Design PPMS Dynacool at temperatures ranging from 10 K to 280 K with $360\degree$ in-plane sample rotation. An external field of 0.3 T was applied to ensure the magnetic moments were fully saturated in-plane (see Section \ref{sup:characterization}). A nonlocal configuration was used where a Keithley 6221 ac source applied a 0.5 mA ac current with a frequency $\omega/(2\pi) =$ 9.8 Hz to one of the Pt strips (injector), and the first- and second-harmonic nonlocal voltage signals were simultaneously measured in the second Pt strip (detector) by two SR-830 lock-in amplifiers with 300 ms time constants. 



\section{Structural and Magnetic Characterization of LAFO Thin Films}
\label{sup:characterization}
\begin{figure*}[!htbp]
   \centering
   {\includegraphics[scale = 0.4]{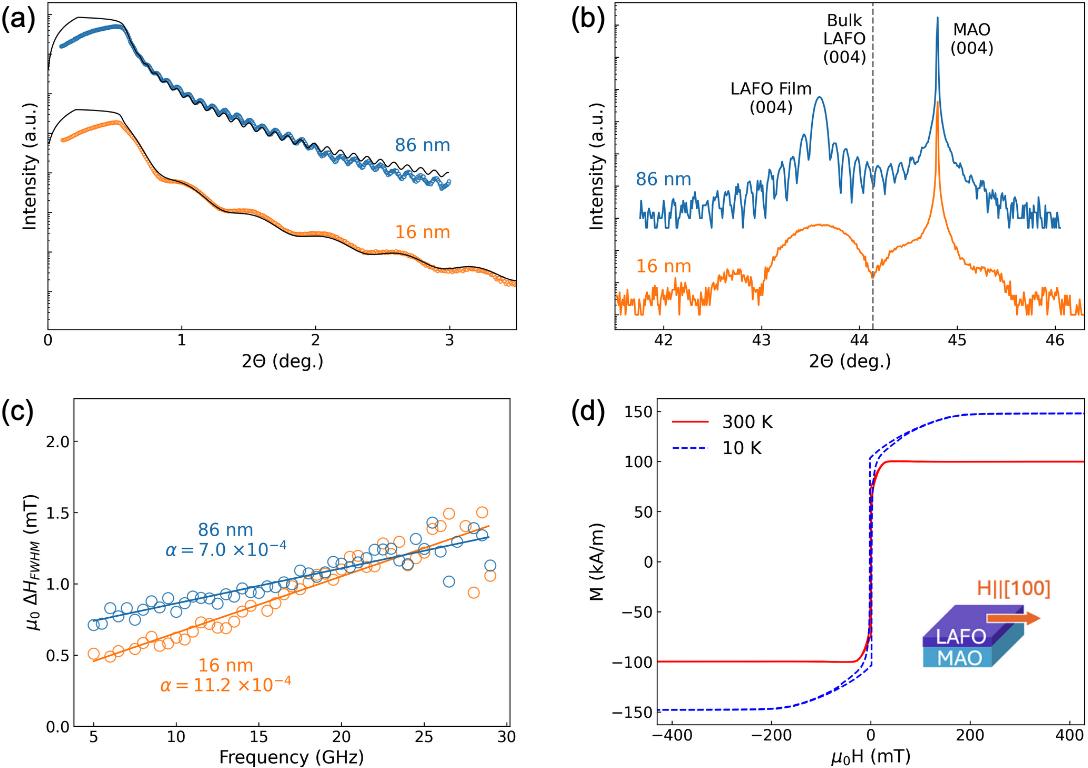}}
   \caption{Structural and magnetic characterization of LAFO thin films. (a) X-ray reflectivity measurement with best fit curves for 16 nm and 86 nm LAFO films. (b) X-ray diffraction measurement for 16 nm and 86 nm LAFO films, showing the LAFO (004) and MAO (004) peaks. (c) Ferromagnetic resonance linewidth as a function of frequency, yielding Gilbert damping parameters of $\alpha \sim 11\times10^{-4}$ and $\alpha \sim 8 \times10^{-4}$ for the 16 nm and 86 nm films, respectively. (d) Magnetic hysteresis loops measured in-plane at 300 K and 10 K, showing temperature dependent in-plane magnetization and saturation fields. }
   \label{fig:characterization}
\end{figure*}

Thickness and crystalline quality were determined using the PANalytical X'Pert 2 in the X-ray reflectivity (XRR) and X-ray diffraction (XRD) configurations, respectively. Figure \ref{fig:characterization}(a) shows the XRR measurement with best fit lines used to determine the LAFO film thickness for 16 nm and 86 nm films. Figure \ref{fig:characterization}(b) shows the XRD spectrum of two epitaxial (001) LAFO films grown on (001) MAO single crystal substrates. We see clear Laue oscillations around the Bragg peak for both film thicknesses, indicating excellent crystallinity. Additionally, the LAFO film peaks are to the left of the bulk LAFO film, indicating the LAFO film is compressively strained to the MAO substrate and has easy in-plane magnetic anisotropy \cite{Omahoney2023}. The Gilbert damping parameter of the samples was measured using ferromagnetic resonance measurements using a coplanar waveguide geometry. Figure \ref{fig:characterization}(c) shows the FMR spectrum of the two epitaxial (001) LAFO films grown on a (001) MAO single crystal substrate. We find the Gilbert damping parameter $\alpha$ by fitting to the slope of the FMR linewidth $\Delta H_{FWHM}$ vs. frequency $f$ data. We find $\alpha \sim 11 \times10^{-4}$ for the 16 nm LAFO film and $\alpha \sim 8 \times10^{-4}$ for the 86 nm film. The static magnetic properties of the samples were measured using the Quantum Design MPMS XL SQUID magnetometer. Figure \ref{fig:characterization}(d) shows the hysteresis loops for the 16 nm LAFO sample at 300 K and 10 K. From the hysteresis loops, we find the saturation field for the films to be $\sim 200$ mT at 300 K and $\sim 0.3$ T at 10 K.  


\section{Derivation of the magnon spin diffusion model} 
\label{sup:SDL equation derivation}
\renewcommand{\theequation}{A\arabic{equation}}
The spin diffusion in thin films is modeled by the one-dimensional diffusion equation given by \cite{dyakonov_1971,valet_1993}
\begin{equation}
    \frac{d^2 n_m}{dx^2} = \frac{n_m}{\lambda_m^2},\,\,\,\,\lambda_m = \sqrt{D\tau_{nc}}, 
    \label{eq:NL full equation}
\end{equation}
where $n_m$ is the magnon density, $D$ is the magnon diffusion constant, $\tau_{nc}$ is the magnon nonconserving relaxation time and $x$ is the direction perpendicular to the Pt electrodes. The 1D approximation is valid in our case because the LAFO thickness (16 nm or 86 nm) is much smaller than the separation distance between the electrodes (1 - 10 $\mu$m), and the electrode length (40 $\mu$m) is much larger than the separation distance. Solutions to Eq.~\eqref{eq:NL full equation} are exponential functions. When combined with boundary conditions that assume there is an initial population at the injector bar $n_m(0)=n_0$ and all magnons at the detector bar a distance $d$ away are absorbed $n_m(d) = 0$, result in the magnon diffusion current density becomes:
\begin{equation}
    j_m \propto \left.\frac{d n_m}{dx}\right|_{x=d} = \frac{\exp(-d/\lambda_m)}{1 - \exp(-2 d /\lambda_m)}. 
    \label{eq:NL full solution}
\end{equation}

\section{Mid-wave infrared microscopy of injector-induced temperature profiles} \label{sup:thermal}
\begin{figure*}[!htbp]
   \centering
   {\includegraphics[scale = 0.4]{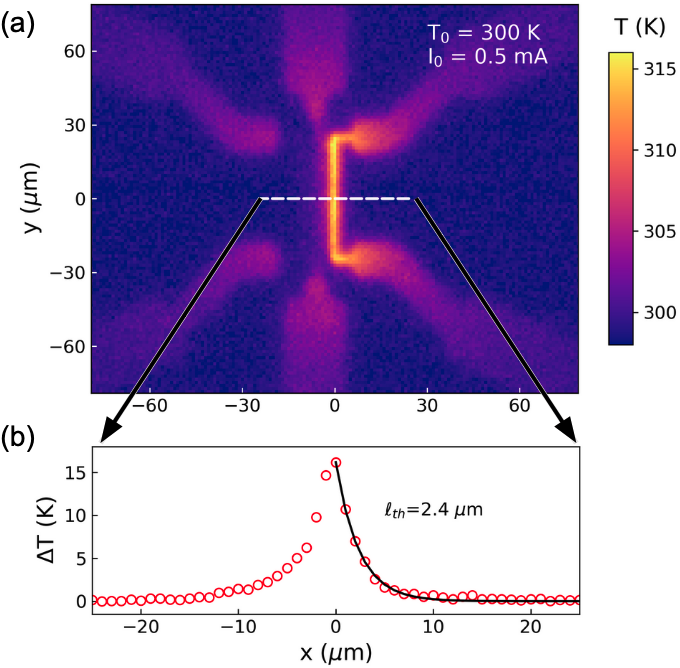}}
   \caption{Thermal characterization of LAFO film with device structure. (a) Mid-wave infrared (MWIR) image of a device ($d=5.0\,\mu$m) under $I_0$ = 0.5 mA at $T_0$ = 300 K, showing the spatial decay of the temperature from the injector bar. (b) Extracted average temperature ($\Delta T = T - T_0$) profile as a function of distance from the injector Pt strip fit, exhibiting exponential decay with characteristic decay length $\ell_{th} =2.4 \,\mu$m. This decay length sets the lower bound for the fitting range used to extract $\lambda_m^{e,th}$ by excluding regions where the temperature gradients are significant.}
   \label{fig:thermal}
\end{figure*}

Mid-wave infrared (MWIR) microscopy was performed in a Quantum Focused Instruments (QFI) Infrascope operating at room pressure. The temperature was controlled by an integrated thermal stage from the manufacturer. Figure~\ref{fig:thermal}(a) is a MWIR microscopy image of a device with Pt bar width w = 0.2 $\mu$m and a separation distance $d = 5 \,\mu$m at $T_0=$ 300 K, with an ac current $I_0=$ 0.5 mA at the injector. Figure~\ref{fig:thermal}(b) shows the spatial decay of the temperature ($\Delta T = T - T_0$) along the $x$-direction, perpendicular to the length of the Pt bar with $x=0\,\mu$m at the center of the Pt bar, that is averaged over the full length L = 40 $\mu$m of the Pt bar. We find a thermal decay length, $\ell_{th}$, of $\sim 2.4 \,\mu$m from fitting an exponential to the temperature decay from the center of the injector bar (the black solid line). 

\section{Temperature and Device Distance Dependence of Nonlocal Resistance}
\label{sup:RvsT diff devs}
In order to determine the spin diffusion length associated with electrically and thermally generated magnons, we measured the nonlocal resistance amplitude for devices with different separation distances $d$ between the two Pt electrodes at different temperatures. The temperature dependence of the nonlocal resistance amplitude is shown at several different separation distances in Fig.~\ref{fig:data}(a, b) for the electrically and thermally excited magnons, respectively. The nonlocal resistance for electrically generated magnons increases with increasing temperature, with an onset around 50 K for shorter separations (red circles) and around 100 K for longer separations (blue triangles). For this reason, we are only able to extract the spin diffusion length for temperatures greater than 90 K for electrically generated magnons. For thermally generated magnons, the nonlocal resistance decreases with increasing temperature, and the spin diffusion length is extracted for all measured temperatures.

\bigskip
\bigskip

\begin{figure*}[!htbp]
   \centering
   {\includegraphics[scale = 0.3]{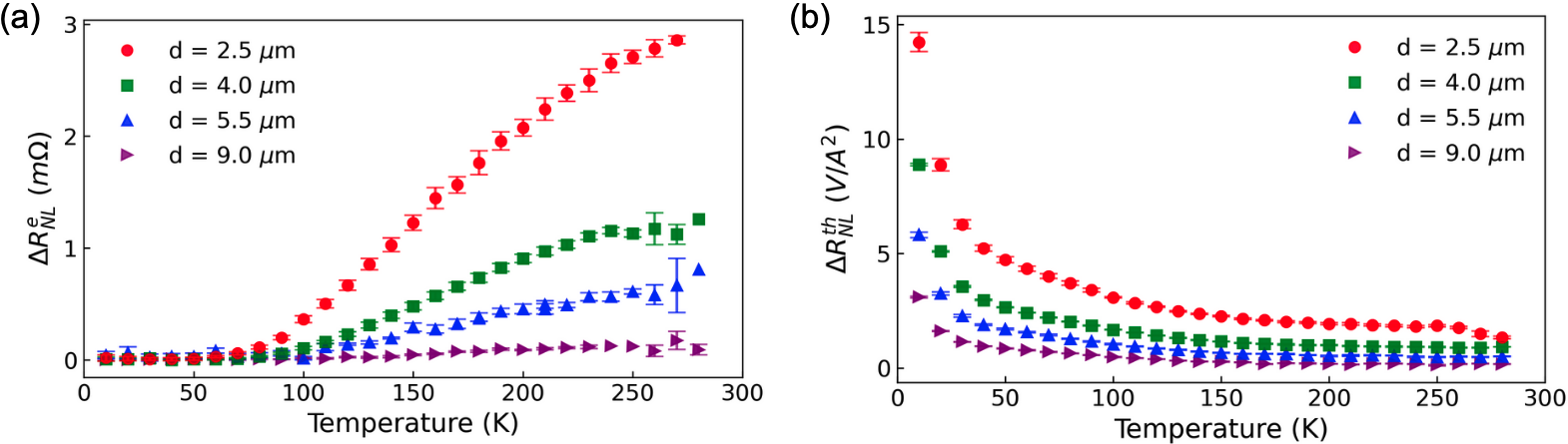}}
   \caption{Nonlocal resistance amplitude as a function of temperature for (a) electrically and (b) thermally generated magnons for different separation distances $d$ between injector and detector Pt strips. For long separation distances, the electrical nonlocal resistance amplitude drops to zero around 100 K, whereas for shorter distances the amplitude reaches zero around 50 K.}
   \label{fig:R vs T diff d}
\end{figure*}

\section{Theoretical Model for magnon diffusion and relaxation mechanisms} 
\label{sup:theory}
\setcounter{equation}{0}
\renewcommand{\theequation}{B\arabic{equation}}

The magnon dynamics are governed by the steady-state Boltzmann equation,
\begin{align}
 \bm{v}_{\bm{k}}\cdot\nabla_{\bm{r}} n_{\bm{k}}
= -\frac{n_{\bm{k}}-n^{\mathrm{loc}}_{\bm{k}}}{\tau_{c,\bm{k}}}  -\frac{n_{\bm{k}}-n^{\mathrm{eq}}_{\bm{k}}}{\tau_{nc,\bm{k}}}
\label{eq:E2}
\end{align}
where $\bm{v}_{\bm{k}}=(1/\hbar)\nabla_{\bm{k}}\varepsilon_{\bm{k}}$ is the magnon group velocity. We write the collision integral in the relaxation-time approximation and decompose the relaxations into magnon-number--conserving channels ($\tau_{c}$) and magnon-number--nonconserving channels ($\tau_{nc}$), with a clear hierarchy of time scales,
$\tau_c \ll t \ll \tau_{nc}$. Due to this separation of time scales, fast scattering processes drive the magnon distribution into a local equilibrium $n^{\mathrm{loc}}_{\bm{k}}$ on a time scale $\sim\tau_c$, while slow scattering processes relax the system toward a global equilibrium distribution $n^{\mathrm{eq}}_{\bm{k}}$ on a much longer time scale $\sim\tau_{nc}$.

\begin{figure}[!htbp]
    \centering
    {\includegraphics[width=0.6\linewidth]{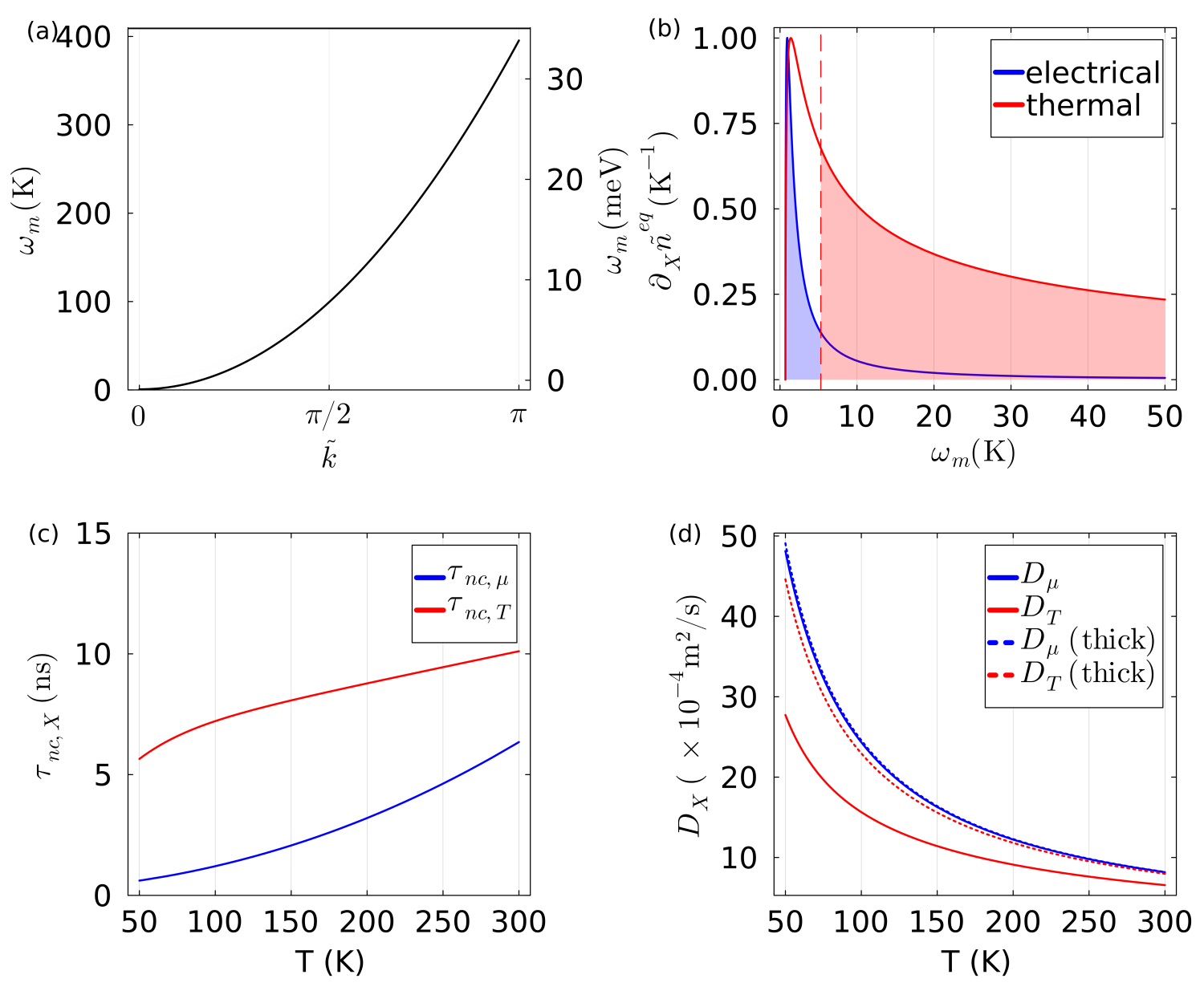}}
    \caption{(a) Magnon dispersion,
    $\hbar\omega_m = k_B \alpha \tilde{k}^2 + k_B \Delta$, where $\tilde{k}= a k$ is the normalized wave vector with the lattice constant of LAFO $a=0.82$ nm, the exchange stiffness is $\alpha=40$~K, and the magnon gap is $\Delta=0.7$~K under $0.3$ T external field \cite{takana_2026}. (b) Non-equilibrium magnon energy distribution at $T=300$ K and $\mu=0$ for electrical and thermal magnons, where the distributions $\partial_X \tilde{n}^{eq}$ are normalized by their maximum value for comparison. Here, we introduce a cutoff energy $\omega_{\mathrm{cut}}=5.3$ K for the upper bound of the electrical magnons and for the lower bound of the thermal magnons. The upper limit of thermal magnon energy is 300 K, which is already close to the zone boundary. The integration regions are shaded. Temperature dependence of (c) the magnon--nonconserving relaxation time $\tau_{nc,X}$ and (d) the diffusion constant $D_X$ for electrical ($X=\mu$) and thermal ($X=T$) magnons. The Rayleigh-type scattering and magnon--phonon scattering parameters are
    $c_1=3~\mathrm{ns}^{-1}$ and $c_2=0.1~\mathrm{ns}^{-1}\,\mathrm{K}^{-1}$, respectively. Dashed lines in $D_X$ indicate the diffusion constant for thicker films, corresponding to a reduced Rayleigh scattering strength $c_1/10$. The TLS exchange splitting is $\omega_{ex}=14$~K, and the intrinsic TLS relaxation time is $\tau_0=48$~ps. The dimensionless TLS scattering strength is $c_3=7.4\times10^{-3}$. The magnon-dispersion parameters $\alpha$ and $\Delta$ and the diffusion-constant parameters $c_1$ and $c_2$ are taken from the Brillouin light-scattering and magnon-conductivity measurements in Ref.~\cite{takana_2026}, while the cutoff energy $\omega_{\mathrm{cut}}$ and the TLS-scattering parameters, $c_3$ and $\tau_0$, are used as fitting parameters. }
    \label{fig:D_tau_nc}
\end{figure}

Taking the zeroth moment of Eq.~(\ref{eq:E2}) yields
\begin{align}
\nabla_{\bm{r}}\cdot \sum_{\bm{k}} \bm{v}_{\bm{k}} n_{\bm{k}}
= - \sum_{\bm{k}} \frac{n_{\bm{k}}-n^{\mathrm{eq}}_{\bm{k}}}{\tau_{nc,\bm{k}}},
\label{eq:E6}
\end{align}
where the contribution from the fast collision channel vanishes due to magnon-number conservation:
\begin{align}
 \sum_{\bm{k}} \frac{n_{\bm{k}}-n^{\mathrm{loc}}_{\bm{k}}}{\tau_{c,\bm{k}}}
= 0.
\label{eq:E7}
\end{align}
We write the nonequilibrium magnon distribution as a small deviation from equilibrium,
\begin{align}
n_{\bm{k}} &= n^{\mathrm{eq}}_{\bm{k}} + \delta n_{\bm{k}}, \\
\delta n_{\bm{k}}  &= \frac{\partial n^{\mathrm{eq}}_{\bm{k}}}{\partial X}\sum_{n=0}^{\infty} g_n(x)\, P_n(\cos \theta),
\label{eq:dnk_legendre}
\end{align}
where $n^{\mathrm{eq}}_{\bm{k}}$ is the equilibrium magnon distribution given by the Bose function,
$n^{\mathrm{eq}}_{\bm{k}}=\left[e^{(\varepsilon_{\bm{k}}-\mu)/k_B T}-1\right]^{-1}$.
We expand $\delta n_{\bm{k}}$ in Legendre polynomials $P_n(\cos\theta)$, appropriate for one-dimensional diffusion, where $\theta$ is the angle between $\bm{k}$ and the diffusion direction~\cite{Fert_1993,zhang_2012}. Here, $X$ denotes the perturbation variable, either the chemical potential ($X=\mu$) or the temperature ($X=T$). Figure~\ref{fig:D_tau_nc}(b) presents the nonequilibrium magnon distributions obtained from the first-order perturbation with respect to the chemical potential and temperature, corresponding to electrically and thermally generated magnons, respectively. 

After straightforward algebra, we obtain the magnon diffusion equation for each type of perturbation:
\begin{align}
D_X \nabla_{\bm{r}}^2 \delta n_m
=
\frac{\delta n_m}{\tau_{nc,X}},
\label{eq:E12}
\end{align}
where $\delta n_m$ is the excess magnon number density. The corresponding diffusion constant and magnon-number--nonconserving relaxation rate are
\begin{align}
\label{eq:E9}
D_X
&=
\left[
\int d \omega_k g(\omega_k)
\frac{1}{3}\tau_{c,k} v_{k}^2
\frac{\partial n^{\mathrm{eq}}_k}{\partial X}
\right]
\left(
\frac{\partial n^{\mathrm{eq}}_m}{\partial X}
\right)^{-1}. \\
\tau^{-1}_{nc,X}
&=
\left[
\int d \omega_k g(\omega_k)
\frac{1}{\tau_{nc,k}}
\frac{\partial n^{\mathrm{eq}}_k}{\partial X}
\right]
\left(
\frac{\partial n^{\mathrm{eq}}_m}{\partial X}
\right)^{-1},
\label{eq:E11}
\end{align}
where $g(\omega_k)$ is the magnon density of states and $n_m=\int d\omega_k\, g(\omega_k)\, n^{\mathrm{eq}}_{k}$ is the equilibrium magnon number density. Note that we replace the wave vector $\bm{k}$ by its magnitude $k$, assuming an isotropic magnon dispersion.

The corresponding magnon diffusion lengths are defined as
\begin{equation}
\lambda_m^X=\sqrt{D_X\tau_{nc,X}},
\label{eq:E13}
\end{equation}
for electrically generated magnons $X=\mu$ ($\lambda_m^{\mu}\equiv \lambda_m^{e}$) and thermally generated magnons $X=T$ ($\lambda_m^T\equiv \lambda_m^{th}$).

The magnon-number--conserving relaxation rate entering the diffusion constant
[Eq.~\eqref{eq:E9}] is modeled as the sum of two contributions,
\begin{align}
\tau^{-1}_{c,\tilde{k}}
= c_1 \tilde{k}^{4} + c_2 \tilde{k}^2 T ,
\label{eq:E14}
\end{align}
where the $\tilde{k}^{4}$ term represents Rayleigh-type scattering induced by
bulk defects or surface roughness, while the second term describes
anisotropy-mediated magnon--phonon scattering~\cite{Streib2019,Liu2017}.
Here $\tilde{k}=ak$ is the dimensionless wave vector normalized by the lattice
constant. 
The coefficients $c_1$ and $c_2$ have units of
$\mathrm{ns}^{-1}$ and $\mathrm{ns}^{-1}\,\mathrm{K}^{-1}$, respectively.

The magnon-number--nonconserving relaxation rate is modeled by a relaxational
mechanism due to two-level systems (TLS)~\cite{VVleck1963,Seiden1964},
\begin{align}
\tau^{-1}_{nc,\tilde{k}}
= c_3 \omega_{ex}
\frac{\omega_m \tau_{\mathrm{imp}}(T)}
{1+\omega_m^2 \tau_{\mathrm{imp}}^2(T)}
\frac{\hbar \omega_{ex}}{k_B T}
\cosh^{-2}\!\left(\frac{\hbar \omega_{ex}}{2 k_B T}\right),
\label{eq:E15}
\end{align}
where $c_3$ is a dimensionless parameter and $\omega_{ex}$ denotes the exchange
coupling between the TLS and the host spins.
The TLS relaxation time is temperature dependent and given by
\begin{equation}
\tau_{\mathrm{imp}}(T)
= \tau_0 \tanh\!\left(\frac{\hbar \omega_{ex}}{2 k_B T}\right).
\label{eq:E16}
\end{equation}

Figure~\ref{fig:D_tau_nc} (c,d) shows the temperature dependence of the magnon--nonconserving relaxation time $\tau_{nc}$ and the diffusion constant $D$ for thermally and electrically generated magnons. The magnon--nonconserving relaxation time $\tau_{nc}$ increases with increasing temperature, reflecting the reduced efficiency of TLS scattering when the two levels become nearly equally populated at high temperatures. In contrast, the diffusion constant $D$ decreases with increasing temperature, as the magnon-number--conserving relaxation is dominated by magnon--phonon scattering whose contribution increases with increasing temperature. We note that the difference between the diffusion constants for thermal ($D_T$) and electrical ($D_\mu$) magnons originates primarily from the
Rayleigh-type scattering, which is more effective for high-$k$ magnons. 

Moreover, since the magnon distributions entering the transport coefficients are weighted by $\partial n_{\bm{k}}/\partial X$ with $X=T,\mu$, it follows that $\tau_{nc,T}$ increases with temperature more slowly than $\tau_{nc,\mu}$. Consequently, the temperature dependence of the thermal magnon diffusion length is overwhelmed by magnon--phonon scattering, leading to a monotonic decrease with increasing temperature, see Fig.~\ref{fig:SDL vs T w/ theory}.

Tuning the film thickness effectively controls the relative strength of Rayleigh-type scattering in the magnon-number--conserving relaxation processes. For thicker films, reduced surface scattering leads to a suppression of the Rayleigh-type contribution. However, since Rayleigh-type scattering scales as $\tilde{k}^4$, it is particularly sensitive to high-$k$ magnons. As a consequence, a pronounced change in the diffusion constant is observed for thermal magnons ($D_T$), whereas the variation of $D_\mu$ remains marginal, see Fig.~\ref{fig:D_tau_nc} (d).

The cutoff energy separates electrically and thermally generated magnons into low-$k$ and high-$k$ sectors. This distinction reflects the higher characteristic energy scale of thermal magnons relative to electrical magnons. The low-$k$ nature of the electrical magnons can be understood from the magnon injection process at the Pt/LAFO interface, which is modeled as direct spin transfer via interfacial $s$-$d$ coupling~\cite{zhang_2012,Bender_2015}. The conversion efficiency is resonantly enhanced when the spin chemical potential, \textit{i.e.}, the chemical potential difference between spin-up and spin-down itinerant electrons in Pt, matches the magnon energy. The spin chemical potential accumulated at the Pt/LAFO interface is generated by the spin Hall effect in Pt~\cite{cornelissen_2016_2},
\begin{align}
\mu_s = 2 \theta_{\mathrm{SH}} J_c \rho_{\mathrm{Pt}} \lambda_s \tanh\left(\frac{t}{2\lambda_s}\right),
\end{align}
where $\theta_{\mathrm{SH}}$, $\lambda_s$, and $\rho_{\mathrm{Pt}}$ denote the spin Hall angle, spin diffusion length, and resistivity of Pt, respectively. Here, $J_c$ is the charge current density and $t$ is the thickness of the Pt strip. In our device, the applied current is $0.5$ mA in the ac measurement, corresponding to $0.35$ mA in dc. The Pt strip has dimensions of $4$ nm in thickness, $200$ nm in width, and $40~\mu\mathrm{m}$ in length. Its resistance is approximately $30~\mathrm{k}\Omega$. Using $\theta_{\mathrm{SH}} \approx 0.1$ and $\lambda_s \approx 2$ nm~\cite{Sinova_2015}, we estimate the spin chemical potential at the Pt/LAFO interface to be $\mu_s\approx 80$ $\mu$eV. 

Additionally, the separation of magnon modes into low-$k$ and high-$k$ sectors may indicate that the nonequilibrium magnon distribution cannot be fully captured by a first-order perturbation of the Bose distribution.
We note, however, that the present formalism remains valid if $\partial n^{\mathrm{eq}}_{\bm{k}}/\partial X$ is replaced by another isotropic distribution in momentum space in Eq.~\eqref{eq:dnk_legendre}. But determining the detailed nonequilibrium magnon distribution requires further investigation.

\section{Comparison of exponential and full diffusion model fits}
\label{sup:SDL fit comparison}
\begin{figure*}[!htbp]
    \centering
    {\includegraphics[scale = 0.4]{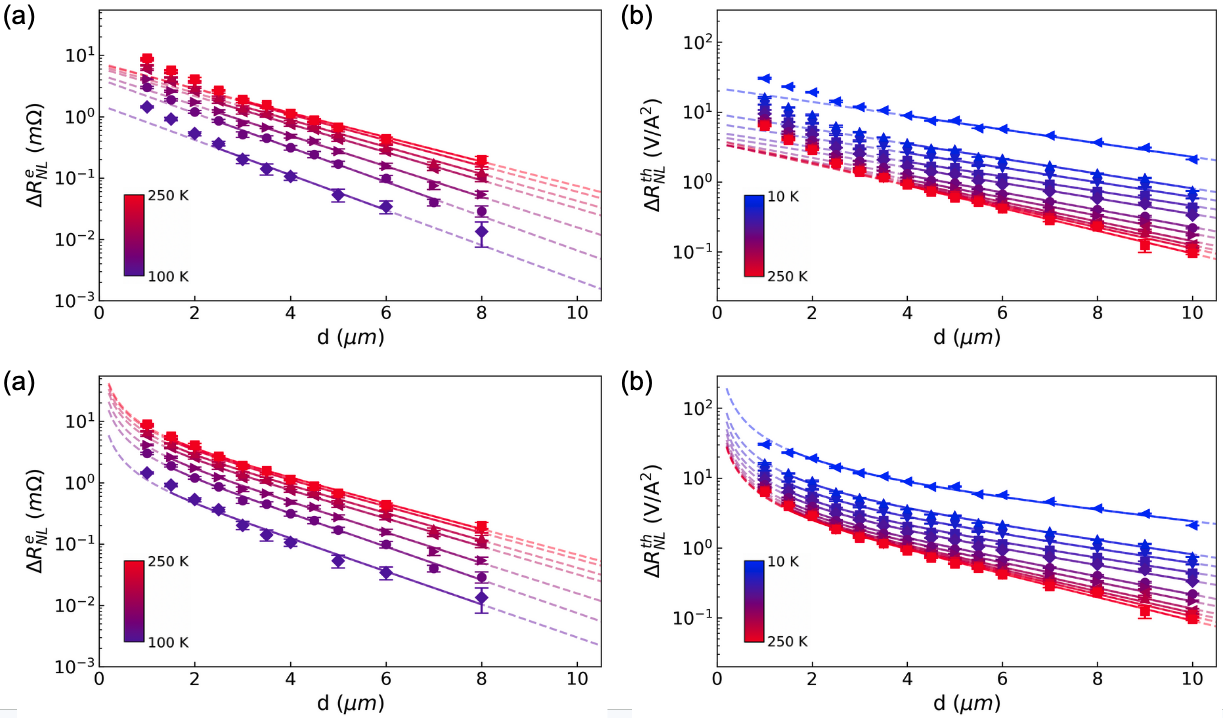}}
    \caption{Comparison between fits to the (a, c) electrical and (b, d) thermal nonlocal voltage amplitude to the full diffusion solution and the exponential form used in the main text. The solid lines in (a,b) are fits to Eq.~\eqref{eq:NL exp sol}, and in (c,d) are fits to Eq.~\eqref{eq:NL full solution}. Both approaches fit well to the data and yield consistent $\lambda_m{e,th}$.}
    \label{fig:data equation comparison}
\end{figure*}

\begin{figure}[!htbp]
    \centering
    {\includegraphics[scale = 0.6]{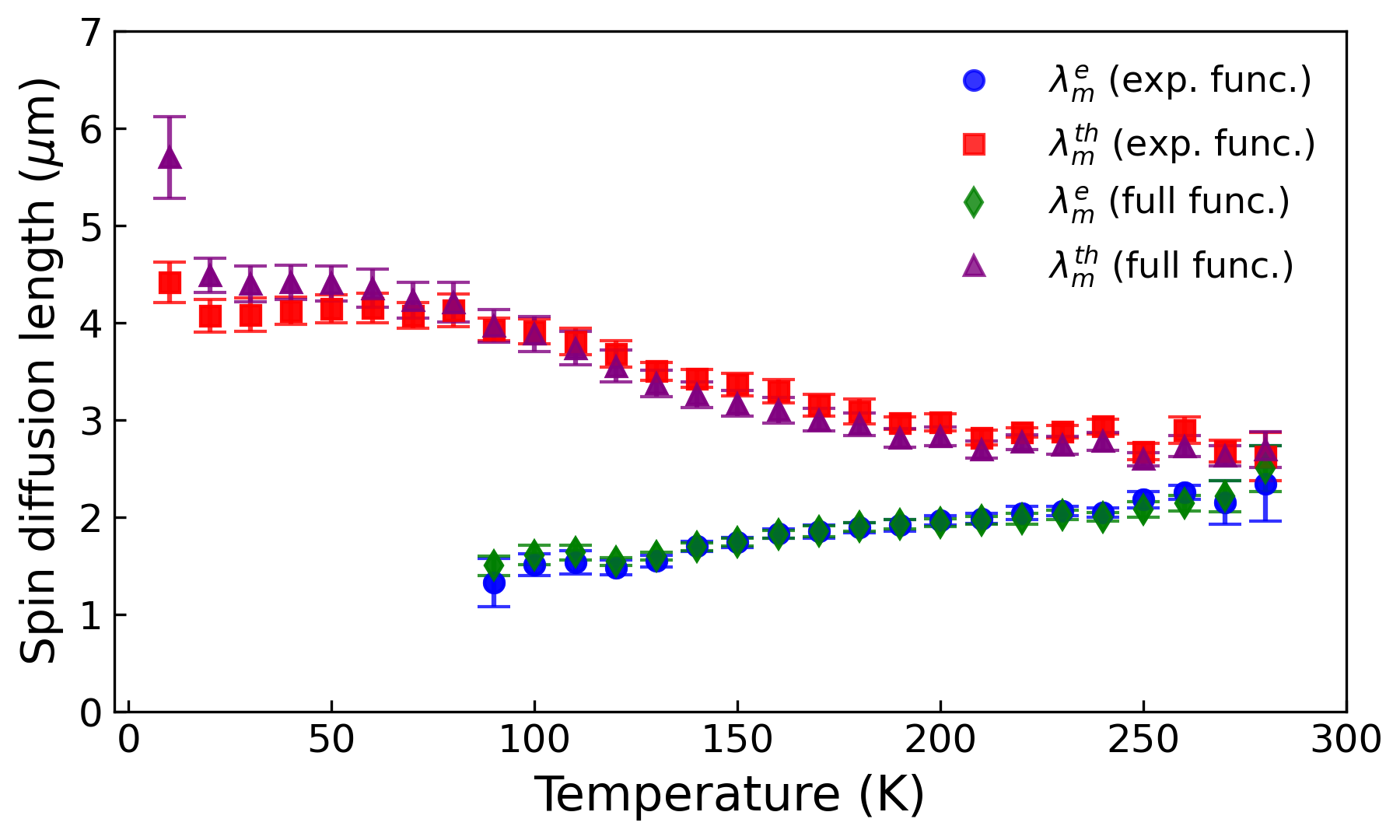}}
    \caption{Spin diffusion length as a function of temperature for thermally ($\lambda_m^{th}$) and electrically generated magnons ($\lambda_m^{e}$), comparing the spin diffusion length extracted using an Eq.~\eqref{eq:NL exp sol} (red and blue) and Eq.~\eqref{eq:NL full solution} (purple and green). The extracted spin diffusion lengths show good agreement for the electrically generated magnons for all temperatures and for the thermally generated magnons, justifying the use of the exponential approximation.}
    \label{fig:SDL equ comparison}
\end{figure}

When fitting for the spin diffusion length, previous studies have used either the full solution (Eq.~\eqref{eq:NL full solution}) or the exponential (Eq.~\eqref{eq:NL exp sol}) with longer distance devices. Figure \ref{fig:data equation comparison} shows the dependence of the nonlocal resistance on the separation of the injector and detector for (a, c) electrically and (b, d) thermally generated magnons when fit to (a, b) the exponential form and (c, d) the full equation. Figure \ref{fig:SDL equ comparison} compares the extracted spin diffusion length values from the two different fit equations. We find no significant difference between the extracted spin diffusion length values when comparing the two different fitting equations. However, it has been argued that using the exponential form provides a more accurate spin diffusion length for thermally generated magnons \cite{cornelissen_2017}, so we report our extracted spin diffusion length from the exponential fit in the main text.

\bigskip
\bigskip

\clearpage

\section{Thickness dependence of LAFO spin diffusion length}
\label{sup:86 nm LAFO data}
\begin{figure*}[!htbp]
    \centering
    {\includegraphics[scale = 0.4]{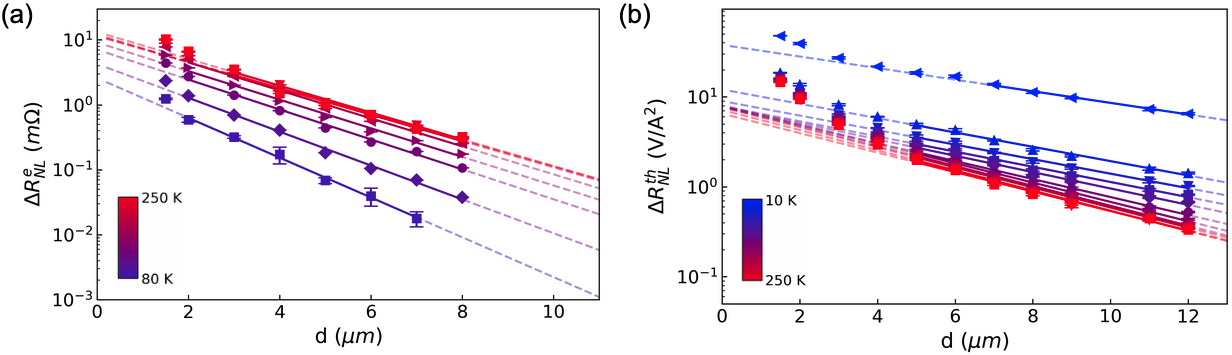}}
    \caption{Nonlocal voltage amplitude for (a) electrically and (b) thermally generated magnons as a function of distance for an 86 nm thick LAFO film. The solid lines are fits to Eq.~\eqref{eq:NL exp sol} and are used to extract the spin diffusion length. }
    \label{fig:data LFO82}
\end{figure*}

\begin{figure}[!htbp]
    \centering
    {\includegraphics[scale = 0.55]{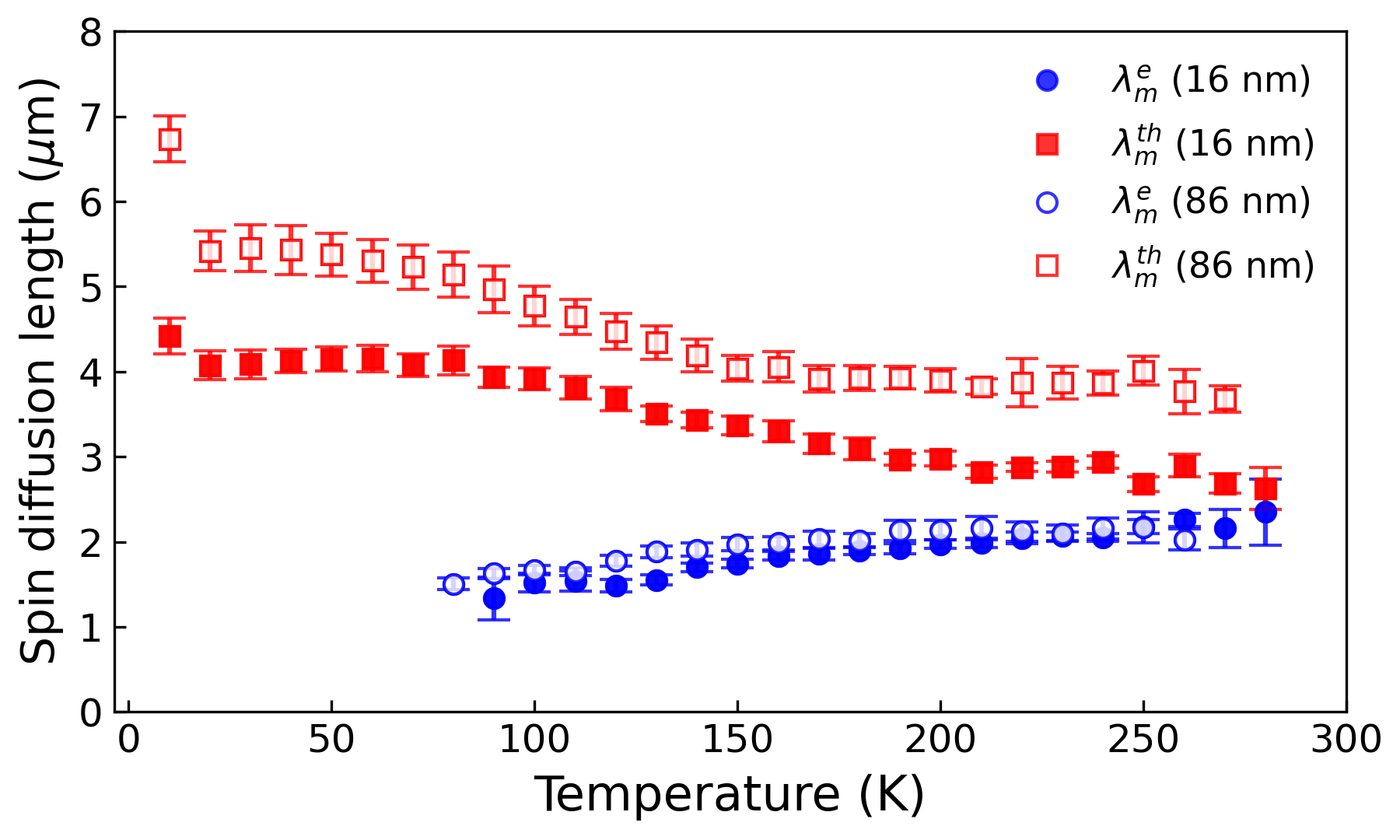}}
    \caption{Spin diffusion length as a function of temperature for thermally ($\lambda_m^{th}$) and electrically generated magnons ($\lambda_m^{e}$), compared for two different thicknesses of LAFO films. Solid shapes denote extracted spin diffusion lengths from a 16 nm LAFO sample, and open shapes are from an 86 nm LAFO sample. For electrically generated magnons, the temperature dependence does not appear to have a thickness dependence. For thermally generated magnons, the trends appear similar for the two LAFO thicknesses, but the thicker LAFO sample has overall larger spin diffusion lengths. The error bars represent the standard statistical error from the least-squares fit.}
    \label{fig:SDL vs T two thicknesses}
\end{figure}

To study the thickness dependence of our LAFO spin diffusion length temperature dependence, we repeat our measurements on a thicker 86 nm film. Figure \ref{fig:data LFO82} shows the (a) electrical and (b) thermal nonlocal resistance data for different separation distances d. The solid lines represent fits to Eq.~\eqref{eq:NL exp sol} where the dashed lines extend the fit to distances not used in the fit. Figure \ref{fig:SDL vs T two thicknesses} compares the extracted spin diffusion length temperature dependence of 16 nm LAFO (filled shapes) with 86 nm LAFO (open shapes). We find that the temperature dependence for electrical and thermal spin diffusion length is similar, with the electrical spin diffusion length values almost exactly the same and the thermal spin diffusion length values overall larger for the thicker LAFO film.

\bigskip
\bigskip

\end{document}